\documentclass{Interspeech}
\usepackage{makecell}
\usepackage{xcolor}
\usepackage{graphicx}
\usepackage{multirow}
\usepackage[colorinlistoftodos]{todonotes}
\usepackage{tipa}
\usepackage{lmodern}

\usepackage{tipa}
\usepackage{arabtex}
\usepackage{utf8}
\setcode{utf8}

\interspeechcameraready

\title{Towards a Unified Benchmark for Arabic Pronunciation Assessment: Qur’anic Recitation as Case Study}

\author[]{%
  Yassine El Kheir$^{*,1}$, 
  Omnia Ibrahim$^{*,2}$, 
  Amit Meghanani$^{*,3}$, 
  Nada Almarwani$^{4}$, 
  Hawau Olamide Toyin$^{5}$, 
  Sadeen Alharbi$^{6}$, 
  Modar Alfadly$^{6}$, 
  Lamya Alkanhal$^{6}$, 
  Ibrahim Selim$^{7}$, 
  Shehab Elbatal$^{7}$, 
  Salima Mdhaffar$^{8}$, 
  Thomas Hain$^{3}$, 
  Yasser Hifny$^{7}$, 
  Mostafa Shahin$^{9}$, 
  Ahmed Ali$^{6}$
}

\affiliation{}{%
  DFKI;  
  $^2$Alexandria Univ.;  
  $^3$Univ. of Sheffield;  
  $^4$Taibah Univ.;  
  $^5$MBZUAI;  
  $^6$SDAIA;  
  $^7$Helwan Univ.;  
  $^8$Univ. of Avignon;  
  $^9$Univ. of New South Wales
}{}

\email{
yassine.el\_kheir@dfki.de, \* Equal Contribution,
}

\begin{document}

\maketitle

\keywords{Qur'an pronunciation, speech recognition, dataset, baseline system, computational paralinguistics}

\begin{abstract} We present a unified benchmark for mispronunciation detection in Modern Standard Arabic (MSA) using Qur'anic recitation as a case study. Our approach lays the groundwork for advancing Arabic pronunciation assessment by providing a comprehensive pipeline that spans data processing, the development of a specialized phoneme set tailored to the nuances of MSA pronunciation, and the creation of the first publicly available test set for this task, which we term as the Qur'anic Mispronunciation Benchmark (QuranMB.v1). Furthermore, we evaluate several baseline models to provide initial performance insights, thereby highlighting both the promise and the challenges inherent in assessing MSA pronunciation. By establishing this standardized framework, we aim to foster further research and development in pronunciation assessment in Arabic language technology and related applications. All models and datasets are available at: \url{https://huggingface.co/IqraEval}.

\end{abstract}

\section{Introduction}

Computer‐aided Pronunciation Training (CAPT) has become an essential tool for self-directed language learners by providing real-time feedback through systematic evaluation and correction of pronunciation errors \cite{neri2008effectiveness}. CAPT systems leverage advances in speech technology, curriculum management, and learner assessment to guide learners toward improved pronunciation. A critical component of these systems is Mispronunciation Detection and Diagnosis (MDD), which identifies pronunciation errors to facilitate targeted feedback.
Arabic presents unique challenges for CAPT due to its linguistic complexity and diverse varieties. The language includes Classical Arabic (the language of the Qur'an and the classical literature), Modern Standard Arabic (MSA, used in formal settings such as education, government, and media), and Dialectal Arabic (the language of everyday conversation) \cite{kheir2024beyond}. In this work, our focus is on MSA—a language that, despite its formal status, is typically acquired as a second language by native Dialect-Arabic speakers.

The phonological system of Arabic is intricate, featuring 34 phonemes: 6 vowels (with distinct short and long forms) and 28 consonants. Among these, the pharyngeal and emphatic consonants play a crucial role \cite{al2004phonetic}. Even minor mispronunciations, such as substituting or omitting emphatic consonants or vowels (e.g., /S/ vs. /s/), can alter semantic meaning and pose significant challenges for automated detection systems \cite{alrashoudi2025improving}.
This challenge is particularly acute in the context of Qur'anic recitation. Governed by strict Tajweed (recitation) rules, precise pronunciation is essential because even slight deviations can distort the intended meaning \cite{balula2021automatic}.

CAPT systems have employed ASR to derive goodness-of-pronunciation (GOP) scores \cite{witt2000phone,kanters2009goodness}. More recent methods have turned to deep learning, utilizing both end-to-end architectures and cascaded pipelines that incorporate pre-trained models \cite{dhleima2024multitask}. In particular, Connectionist Temporal Classification (CTC)-based models integrated with self-supervised acoustic encoders (e.g., wav2vec2.0) have demonstrated notable improvements in mispronunciation detection \cite{wav2vec2, alrashoudi2025improving}.
Despite promising progress in languages like English, adapting these CAPT technologies to Arabic remains challenging.

Arabic Automatic Speech Recognition (ASR) systems omit diacritics—vital markers for phonetic details like vowel length and stress—which further complicates the task \cite{ali2016mgb}. Prior studies have explored various approaches for Arabic MDD—including handcrafted features, CNN-based methods, transfer learning, and transformer-based models \cite{ alrashoudi2025improving, ccalik2024novel}—with several focusing on Qur'anic recitation for both general users and specific groups such as children \cite{alsahafi2024empirical, rahman2021development, harere2023quran}.

However, the major obstacle persists: the absence of standardized resources and benchmarks for Arabic CAPT, compounded by data scarcity and challenges in dialectal standardization. To address these gaps, we present the first open benchmark for mispronunciation detection in MSA with a focus on Qur'anic recitation. Our contributions include:
\begin{itemize}
    \item Compiling a standard pipeline with an optimized phoneme set for Qur'anic Arabic; 
    \item Releasing the first publicly available test set for Qur'an MDD \textbf{QuranMB.v1};
    \item Benchmarking several baseline models to reveal performance trends and specific challenges in assessing MSA pronunciation;
    \item Generating a 52-hour synthetic dataset with simulated errors that achieves competitive accuracy relative to wild-collected data.
\end{itemize}

\section{Dataset Curation}


Due to the limited availability of phonetically labeled Arabic speech, 
we employed two complementary strategies for our training corpus. First, we collected an in‐the‐wild dataset of non‐dialectal Arabic speech, assuming that speakers follow standardized vowelization rules, and automatically vowelized the corresponding transcription. Second, we generated a synthetic corpus using Text-to-speech (TTS) models trained on fully vowelized text, ensuring precise speech-to-phoneme alignment. The test set QuranMB.v1 consists of natural speech data that is collected with vowelization, following our detailed collection guidelines presented later in this section.

\subsection{Training Set: CMV-Ar}

Our training corpus, CMV-Ar, incorporates an $82.37$ hours subset of the Common Voice Dataset \cite{ardila2019common} version $12.0$ specifically for MSA Arabic speech recognition. This dataset consists of read speech samples collected from a diverse pool of speakers with a well-balanced gender distribution. We assume that the provided speech aligns with linguistically driven transcript vowelization; to ensure accuracy, we applied our in-house state-of-the-art vowelizer to the transcriptions, thereby generating fully vowelized transcriptions. Additionally, we augmented the corpus with samples drawn from Qur’anic recitations, where the vowelization is correct\footnote{Assuming that speech and transcript are referring to the same Qirā’ah (way of recitation) from the seven-Qirā’āt} by design. The complete CMV-Ar dataset, including the Qur’an samples, will be made publicly available.

\subsection{TTS Augmentation}

\begin{figure} 
\centering 
\hspace{-0.7cm} 
\includegraphics[width=0.5\textwidth]{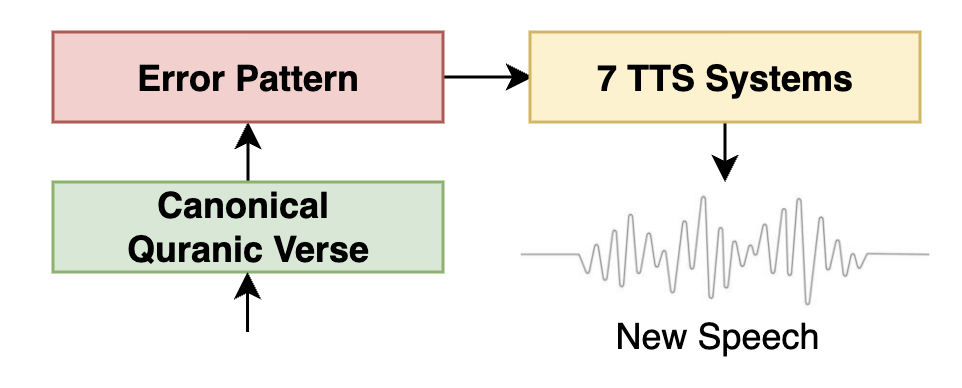}
\caption{TTS Augmentation Pipeline} 
\label{fig:TTS} 
\end{figure}

One of the key challenges in developing reliable mispronunciation detection and diagnosis (MDD) models is the scarcity of annotated mispronounced speech data. A generative speech model capable of mimicking mispronounced speech—demonstrated effectively in \cite{korzekwa2022computer}—could significantly boost our training corpus.

In our CMV-Ar dataset, we assume that the speech perfectly matches the vowelized transcriptions. In practice, however, Arabic speakers often deviate from linguistically driven vowelization by dropping or altering vowels. Furthermore, while we assumed that no character-level mispronunciations occur, Arabic speaker's native dialect influences on MSA articulation can introduce errors \cite{kheir2024beyond}.

To address these limitations, we employed seven in-house single-speaker TTS systems ($5$ male and $2$ female voices) trained on fully vowelized transcriptions to generate canonical speech from a given vowelized text. This process augmented our training dataset with an additional $26$ hours of error-free speech. 

Additionally, as illustrated in Figure \ref{fig:TTS}, we then simulate mispronunciation patterns—particularly those relevant to Qur’anic recitation—by systematically modifying the canonical transcript inputs to incorporate common pronunciation errors. Given a canonical transcript, we randomly select four characters and/or diacritics and modify them based on a predefined confusion pairs matrix. The construction of this matrix is specified in Section \ref{error_pattern}. The resulting transcription— containing injected error patterns—is fed into the TTS system to generate mispronounced speech. In this process, we retain full control over the generated errors, providing complete annotation of the modified patterns. 

The TTS systems generate further a $26$ hours of speech from these erroneous transcriptions, ensuring that a portion of the synthetic data reflects controlled mispronunciation patterns. The full generated TTS ($26$ + $26$) $52$ hours dataset will be publicly available.

\subsection{Testing Dataset: QuranMB.v1}
\begin{figure}
   \centering
    \includegraphics[width=0.6\linewidth]{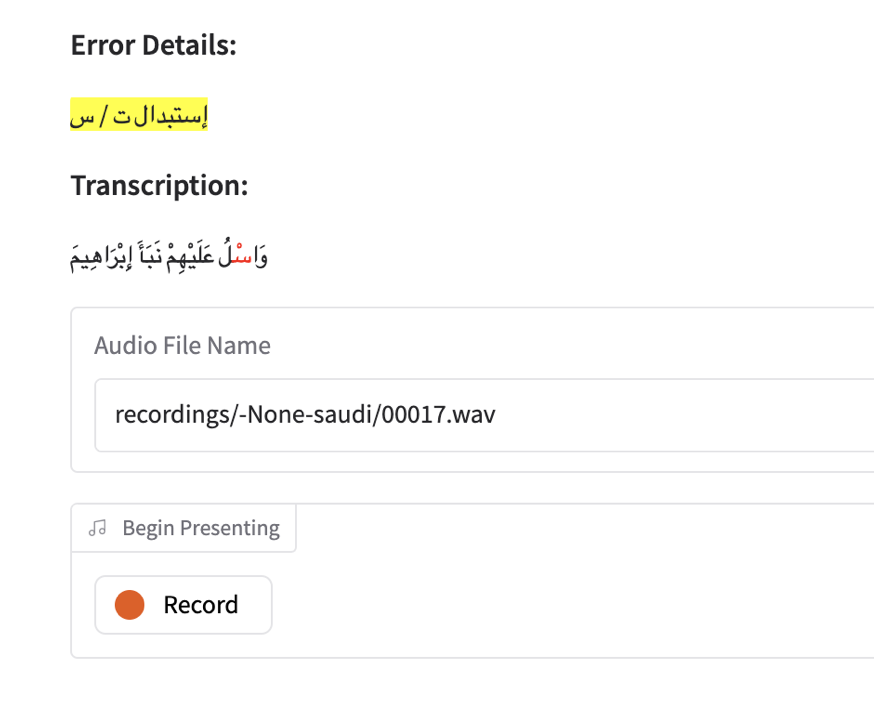}
    \caption{Interface of the recording tool, highlighting modified text and providing instructions to ensure consistent pronunciation errors.}
    \label{fig:recording}
\end{figure}

To construct a reliable test set, we select $98$ verses from the Qur’an, which are read aloud by $18$ native Arabic speakers ($14$ females, $4$ males), resulting in approximately $2.2$ hours of recorded speech. The speakers were instructed to read the text in MSA at their normal tempo, disregarding Qur’anic tajweed rules, while deliberately producing the specified pronunciation errors. To ensure consistency in error production, we develop a custom recording tool that highlighted the modified text and displayed additional instructions specifying the type of error (Figure \ref{fig:recording}). Before recording, speakers were required to silently read each sentence to familiarize themselves with the intended errors before reading them aloud.

\subsection{Error Pattern Construction}
\label{error_pattern}

To systematically introduce pronunciation-based errors, we construct a confusion matrix that maps each character or diacritic to a set of commonly mispronounced counterparts, including deletions. These substitutions simulate realistic phonetic errors by replacing the target character or diacritic with a randomly selected confusable sound. The confusion matrix is derived from phoneme similarity data extracted from \cite{kheir2022speechblender}, ensuring that substitutions align with natural mispronunciation tendencies. For instance, commonly confused phoneme pairs include (\<ت>\  
 /t/ vs. \<ط>\ /T/ ), (\<س>\ /s/ vs. \<ص>\ /S/ ), and (\<غ>\ /g/ vs. \<خ>\ /x/). The confusion dictionary will be publicly available.

\subsection{Data Processing (Describe Phoneme Set)}

Since the goal of this work is to detect mispronounced sounds in Qur’an recitation, the model is designed to recognize the sequence of pronounced phonemes and compare it to the target phoneme sequence.
To accomplish this, correct phoneme transcriptions of the dataset are required as training targets for the model. As our focus on the MSA reading of Qur'an without tajweed rules we employed the phonetizer introduced by Nawar Halabi \cite{halabi2016phonetic} developed for vowelized MSA. This phonetizer was optimized for phonetic coverage in speech synthesis. It employs a greedy algorithm to minimize the size of the speech corpus while maintaining comprehensive phonetic and prosodic coverage. The phonetic vocabulary includes diphones and accounts for features such as stress, pausing, intonation, gemination, and emphaticness. It also addresses specific challenges in MSA, including the influence of diacritics and the need for consistent pronunciation.

Since the phonetizer was designed for speech synthesis, it distinguishes between vowels following emphatic and non-emphatic consonants. However, in MSA, this distinction represents different realizations that do not affect meaning or intelligibility. As our focus is on the MSA style, we disregarded this distinction and treated vowels following emphatic and non-emphatic consonants as the same sound.
A comprehensive inventory of all 68 phonemes, along with detailed descriptions, is available\footnote{\scriptsize\url{https://huggingface.co/spaces/IqraEval/ArabicPhoneme}}. Gemination, the doubling of a consonant sound, is explicitly handled by introducing a phoneme where the consonant symbol is duplicated. For instance, the geminated /b/ is represented as /bb/. This results in a total of 68 phonemes. 

As mentioned earlier, CMV-Ar is vowelized using an in-house vowelizer. For the TTS dataset and QuranMB.v1 test set, the transcription is fully vowelized by design. To obtain correct phoneme sequences, we applied the phonetizer to the vowelized transcription of all these datasets. 

\begin{figure}
\centering
\includegraphics[width=0.5\textwidth]{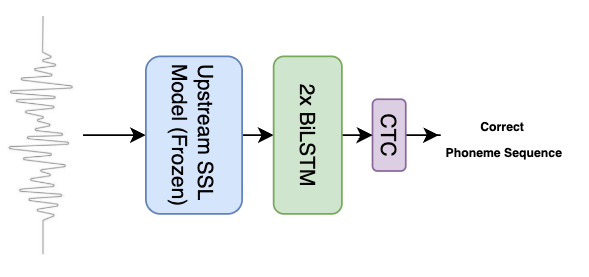}
\caption{Mispronunciation Detection Modeling Pipeline}
\label{fig:modelCTC}
\end{figure}

\section{Methodology}
\label{sec:methodology}

\begin{table}[ht]
    \centering
    \caption{Data configurations for real, synthetic, and combined speech for model training and evaluation.}
    \resizebox{\columnwidth}{!}{%
    \setlength{\tabcolsep}{4pt}
    \begin{tabular}{l p{4.5cm} l}
        \hline
        \textbf{Split} & \textbf{Description} & \textbf{Data Type} \\
        \hline
        \textbf{CMV-Ar} & train set: 79h (71,391 utt.), dev set: 3.33h (2,588 utt.) & Real \\
        \textbf{TTS} & synthetic train set: 45h (47,463 utt.), synthetic dev set adjusted to 3.36h (2,588 utt.) & Synthetic \\
        \textbf{CMV-Ar+TTS} & Combined 79h CMV-Ar with full TTS (52h), resulting in  131h training set, real dev set: 3.33h (2,588 utt.), same as CMV-Ar . & Real + Synthetic \\
        \hline
    \end{tabular}%
    }
    \label{tab:data_splits}

\end{table}

Our framework for MDD leverages self-supervised learning (SSL)-based speech models with subsequent temporal modeling, as depicted in Figure \ref{fig:modelCTC}. 
We have used similar setup as described in the SUPERB \cite{superb} to train the model. In SUPERB, the SSL models weights are frozen. The features obtained from the weighted sum over transformer layers are feeded as input to the model head. The model head consists of a 2-
layer 1024-unit Bi-LSTM network with Connectionist Temporal Classification (CTC) \cite{ctc} loss on phoneme sequence.
To obtain the phoneme sequence during inference, CTC greedy decoding is used. 

\subsection{SSL Model Variants}
We investigate both monolingual and multilingual SSL variants. Our base experiments use 94M-parameter models: English-only Wav2vec2~\cite{wav2vec2}, HuBERT~\cite{hubert}, and WavLM~\cite{wavlm} for monolingual analysis, alongside the multilingual mHuBERT~\cite{mHuBERT} pretrained on 90,430 hours of 147-language data.

\subsection{Data Configuration}
Three data configurations are evaluated to assess model performance using real speech, synthetic speech, and their combination. The details of the splits are shown in Table \ref{tab:data_splits}.
\label{sec:datasets}

\begin{table*}[]
  \centering
  \caption{Experimental Results on QuranMB.v1 test set. TA: True Acceptance, FR: False Rejection, FA: False Acceptance, CD: Correct Diagnosis, ED: Error Diagnosis. $\downarrow$ Lower is better, $\uparrow$ Higher is better.}
  \scalebox{0.8}{
    \begin{tabular}{lcc|ccc|cccc}
      \toprule
                       & \multicolumn{2}{c}{\textbf{canonicals}}                           & \multicolumn{3}{c}{\textbf{mispronunications}}                                                        & & & & \\ 
      \cline{2-10} 
                       & \multicolumn{1}{l}{\textbf{TA (\%)}$\uparrow$} 
                       & \multicolumn{1}{l}{\textbf{FR (\%)}$\downarrow$} 
                       & \multicolumn{1}{l}{\textbf{FA (\%)}$\downarrow$} 
                       & \multicolumn{1}{l}{\textbf{CD (\%)}$\uparrow$} 
                       & \multicolumn{1}{l}{\textbf{ED (\%)}$\downarrow$} 
                       & \multicolumn{1}{l}{\textbf{Recall (\%)}$\uparrow$} 
                       & \multicolumn{1}{l}{\textbf{Precision (\%)}$\uparrow$} 
                       & \multicolumn{1}{l}{\textbf{F1-score (\%)}$\uparrow$} 
                       & \multicolumn{1}{l}{\textbf{PER (\%)}$\downarrow$} \\
      \midrule
      \multicolumn{10}{c}{\textbf{CMV-Ar}} \\ 
      \midrule
      \textbf{Wav2vec2}   & 83.88 & 16.12 & \textbf{23.28} & 61.81 & 38.19 & \textbf{76.72} & 15.71 & 26.08 & 20.17 \\
      \textbf{WavLM}      & 84.27 & 15.73 & 24.65          & 63.00 & 37.00 & 75.35          & 15.80 & 26.12 & 19.74 \\
      \textbf{HuBERT}     & 84.25 & 15.75 & 25.25          & 62.02 & 37.98 & 74.75          & 15.67 & 25.91 & 19.99 \\
      \textbf{mHuBERT}    & \textbf{86.21} & \textbf{13.79} & 24.44 & \textbf{66.78} & \textbf{33.22} & 75.56 & \textbf{17.67} & \textbf{28.64} & \textbf{17.60} \\
      \midrule
      \multicolumn{10}{c}{\textbf{TTS}} \\ 
      \midrule
      \textbf{Wav2vec2}   & 78.94 & 21.06 & 18.98 & 63.02 & 36.98 & 81.02 & 13.09 & 22.54 & 30.55 \\
      \textbf{WavLM}      & 80.43 & 19.57 & 18.37 & 61.81 & 38.19 & 81.63 & 14.04 & 23.96 & 27.44 \\
      \textbf{HuBERT}     & 79.06 & 20.94 & \textbf{17.86} & 64.51 & 35.49 & \textbf{82.14} & 13.31 & 22.91 & 29.78 \\
      \textbf{mHuBERT}    & \textbf{85.46} & \textbf{14.54} & 20.09 & \textbf{70.36} & \textbf{29.64} & 79.91 & \textbf{17.71} & \textbf{29.00} & \textbf{20.32} \\
      \midrule
      \multicolumn{10}{c}{\textbf{CMV-Ar+TTS}} \\ 
      \midrule
      \textbf{Wav2vec2}   & 84.34 & 15.66 & 22.93 & 65.00 & 35.00 & 77.07 & 16.16 & 26.72 & 19.47 \\
      \textbf{WavLM}      & 85.18 & 14.82 & 24.39 & 64.32 & 35.68 & 75.61 & 16.66 & 27.30 & 18.80 \\
      \textbf{HuBERT}     & 84.54 & 15.46 & \textbf{22.57} & 63.99 & 36.01 & \textbf{77.43} & 16.40 & 27.06 & 19.29 \\
      \textbf{mHuBERT}    & \textbf{87.35} & \textbf{12.65} & 25.71 & \textbf{68.26} & \textbf{31.74} & 74.29 & \textbf{18.70} & \textbf{29.88} & \textbf{16.42} \\
      \bottomrule
    \end{tabular}
  }
\end{table*}

\subsection{Training Protocol}
All experiments maintain frozen SSL weights with feature extraction via S3PRL toolkit~\cite{s3prl1,superb}\footnote{\url{https://github.com/s3prl/s3prl}}. We employ Adam optimization with learning rate 1e-4, batch size 16, and early stopping based on phoneme error rate (PER) on the development set of corresponding split. Models train for 12.5k updates, with final selection guided by dev set performance. Final results are reported on our QuranMB.v1 test set.

\subsection{Evaluation Metrics}
\label{sec:evaluation}

Our hierarchical evaluation follows established MDD conventions \cite{leung2019cnn, li2016mispronunciation}, categorizing predictions into four classes: True Accept (TA), True Reject (TR), False Accept (FA), and False Reject (FR), Correct Diagnosis (CD), Error Diagnosis (ED). Precision and Recall derive from diagnostic accuracy:

\begin{equation}
\text{Precision} = \frac{TR}{TR + FR}, \quad 
\text{Recall} = \frac{TR}{TR + FA}
\end{equation}

The primary F1-score combines these through harmonic mean:
\begin{equation}
F_1 = 2\cdot\frac{\text{Precision} \cdot \text{Recall}}{\text{Precision} + \text{Recall}}
\end{equation}

\section{Results and Analysis}
\label{sec:results }

Table \ref{tab:experimental_results} summarizes the performance of SSL-based MDD models across dataset configurations on QuranMB.v1 test set. Key trends emerge in multilingual capability, synthetic data utility, and task complexity.

\subsection{Multilingual vs Monolingual SSL models}
The multilingual mHuBERT model consistently outperforms monolingual SSL variants across all splits, achieving the highest F1-scores (28.64\% on CMV-Ar, 29.00\% on TTS, and 29.88\% on CMV-Ar+TTS). This advantage likely stems from mHuBERT’s pretraining on 147 languages, enabling richer phonetic representations for Arabic mispronunciation detection compared to English-only SSL models. Notably, mHuBERT achieves superior true acceptance (TA) rates (86.21–87.35\%) and lower false rejection (FR) rates (12.65–14.54\%) across configurations, reflecting its robustness to phonetic variations.

\subsection{Synthetic Data Effectiveness} Despite limited speaker diversity, the TTS split demonstrates competitive performance. Specifically, mHuBERT achieves a 29.00\% F1-score with a 29.64\% error diagnosis (ED) rate, outperforming monolingual models trained on CMV-Ar (e.g., HuBERT: 25.91\% F1, 37.98\% ED). This result suggests that synthetic data—by design incorporating mispronunciation patterns—can effectively capture the nuances of recitation errors even with fewer speakers. Moreover, when TTS data is combined with CMV-Ar, mHuBERT's performance further improves, reaching a 29.88\% F1-score, which highlights the complementary benefits of integrating both natural and synthetic datasets.

\subsection{Task Complexity} Establishing initial baselines for Qur'anic pronunciation assessment reveals the considerable complexity of this task. Current models achieve modest performance (F1 $\leq$ 30\%), with precision scores between 13.09\% and 18.70\%. These figures suggest that even advanced multilingual SSL representations find it challenging to capture the subtle articulation errors characteristic of Qur'anic recitation rules, leading to frequent false acceptances. The best-performing model, mHuBERT trained on the CMV-Ar+TTS dataset, achieves a PER of 16.42\%, underscoring the persistent challenges in achieving precise phonetic alignment even with MDD SOTA approaches. These baseline results establish a foundational benchmark for future research, emphasizing the need for novel strategies to address this complexity. Prioritizing data curation—particularly through the inclusion of diverse, real-world mispronunciation patterns—may prove critical in bridging these gaps.

\section{Conclusion}
In this work, we have introduced the first comprehensive public benchmark for MSA pronunciation detection, providing detailed documentation of our data curation methodology, specialized phoneme set, and data augmentation approaches. The release of QuranMB.v1, our test dataset, represents a significant contribution as the first publicly available Qur'anic benchmark specifically designed for pronunciation assessment. Our baseline model evaluations have demonstrated the inherent complexity of this task, highlighting the necessity for carefully curated training data and sophisticated augmentation techniques to effectively capture phonetic variations. Future work will focus on expanding our benchmarking efforts through the collection of diverse datasets, including L2 speakers, which will require extending our phoneme vocabulary to incorporate non-Arabic sounds for more precise assessment. This framework establishes a foundation for advancing Arabic pronunciation assessment technology and opens new avenues for research in this critical domain.

\section{Acknowledgment}
We thank the Saudi Data and Artificial Intelligence Authority (SDAIA) for hosting the Winter School where this work took place, and for the generous computing resources provided.

\bibliographystyle{IEEEtran}
\bibliography{mybib}

\end{document}